\documentclass{article}

\usepackage{PRIMEarxiv}

\usepackage[utf8]{inputenc} 
\usepackage[T1]{fontenc}    
\usepackage{hyperref}       
\usepackage{url}            
\usepackage{booktabs}       
\usepackage{amsfonts}       
\usepackage{nicefrac}       
\usepackage{microtype}      
\usepackage{lipsum}
\usepackage{fancyhdr}       
\usepackage{graphicx}       
\graphicspath{{media/}}     

\usepackage{amsmath}
\usepackage{float}
\newcommand{\pkg}[1]{\textbf{#1}}
\newcommand{\proglang}[1]{\textsf{#1}}

\usepackage{xcolor}
\definecolor{firebrick2}{RGB}{228, 44, 44}
\definecolor{bisque2}{RGB}{238, 200, 171}
\definecolor{lavenderblush3}{RGB}{205, 153, 188}
\definecolor{lightgreen}{RGB}{162, 235, 95}
\definecolor{steelblue}{RGB}{130, 180, 252}
\definecolor{magenta}{RGB}{255, 0, 255}
\definecolor{lightpurple}{RGB}{161, 97, 255}
\definecolor{lightblue}{RGB}{65,65, 225}
\definecolor{brown4}{RGB}{139, 35, 35}
\definecolor{darkgoldenrod3}{RGB}{205, 149, 12}

\title{Visualisations for Exploratory Modelling Analysis of Bayesian Hierarchical Regression Models
\thanks{\textit{\underline{Citation}}: 
\textbf{Authors. Title. Pages.... DOI:000000/11111.}} 
}

\author{
  Oluwayomi Akinfenwa\\
  Hamilton Institute\\
  Maynooth University\\
  Maynooth\\
  Co. Kildare, Ireland \\ 
  \texttt{Oluwayomi.Akinfenwa.2022@mu.ie} \\
   \And
   Niamh Cahill\\
  Department of Mathematics and Statistics\\
  Maynooth University\\
  Maynooth\\
  Co. Kildare, Ireland \\
  \texttt{Niamh.Cahill@mu.ie} \\
  \And
 Catherine Hurley\\
  Department of Mathematics and Statistics\\
  Maynooth University\\
  Maynooth\\
  Co. Kildare, Ireland \\
  \texttt{Catherine.Hurley@mu.ie} \\
}

\begin{document}
\maketitle

\begin{abstract}
When developing Bayesian hierarchical models, selecting the most appropriate hierarchical structure can be a challenging task, and visualisation remains an underutilised tool in this context. In this paper, we consider visualisations for the display of hierarchical models in data space and compare a collection of multiple models via their parameters and hyper-parameter estimates. Specifically, with the aim of aiding model choice, we propose new visualisations to explore how the choice of Bayesian hierarchical modelling structure impacts parameter distributions. The visualisations are designed using a robust set of principles to provide richer comparisons that extend beyond the conventional plots and numerical summaries typically used.  As a case study, we investigate five Bayesian hierarchical models fit using the \pkg{brms} \proglang{R} package, a high-level interface to Stan for Bayesian modelling, to model country mathematics trends from the PISA (Programme for International Student Assessment) database. Our case study demonstrates that by adhering to these principles, researchers can create visualisations that not only help them make more informed choices between Bayesian hierarchical model structures but also enable them to effectively communicate the rationale for those choices.
\end{abstract}


\section{Introduction}\label{sec-intro}

Visualisation has been a longstanding tool within statistical workflows for simplifying complex concepts into comprehensible insights and plays an important role in investigating, understanding and criticising regression models. It also plays a key role in exploratory modelling analysis, which \cite{unwin2018ensemble}, described as ``the evaluation and comparison of many models simultaneously''. 

The importance of visualisation throughout the Bayesian analysis workflow has been well-established by \cite{gabry2019visualization}. They emphasise the benefits of visualisation at all stages, from exploratory data analysis to model development, diagnostics and communication of results. They further explain that visualisation helps sift through networks of increasingly complex models that can capture the features and heterogeneity in the data and they state that visualisation supports model comparison, evaluation and decision-making. 

In the current paper, we are not addressing the entire Bayesian workflow, our goal instead is the design of visualisations for exploring parameters and hyper-parameters across different Bayesian Hierarchical Models (BHMs). To date, little work has been done on visualisation in this context. 

The BHM relies on a statistical modelling procedure that integrates information across many levels, enabling simultaneous estimation of multiple model parameters \cite{gelman2013bayesian}. The model possesses two essential features. Firstly, it has a hierarchical or multi-level structure. Secondly, it uses prior distributions to reflect available information, even if it is vague about the probable values and variability at each hierarchical level \cite{mcglothlin2018bayesian}. The model uncertainties are propagated and quantified by combining the BHM prior structure with the model for the observed data. 

The BHM is particularly valuable in situations where sparse data poses a challenge for parameter estimation \cite{rue2007approximate}. The structure enables the leveraging of strength from group level parameters to enhance estimates at the individual-level, resulting in more reliable and robust estimates. In addition, the inherent shrinkage of parameter estimates, pulling estimates toward the overall population distribution, can reduce the influence of outliers in the data. 

In this paper, via a set of visualisation principles, we specifically explore the importance of visualisation for BHMs for gaining insights into model choice and interpretation of model parameters. We follow the strategies for visualising statistical models presented by \cite{wickham2015visualizing}, which are the display of models in data space and examining a collection of multiple models. Our visualisation designs give careful consideration to choices of layout, colour, ordering and scaling to facilitate effective comparisons.

To demonstrate the relevance of our visualisation designs to a real-life dataset, we analyse a subset of the PISA (Programme for International Student Assessment) dataset, obtained from the OECD (Organisation for Economic Cooperation and Development) website \cite{pisadata}. 
PISA, initiated by the OECD in 1997, provides regular and timely assessments of student achievement with policy-oriented comparable indicators. The surveys, conducted among 15-year-old students, assess performance in reading, mathematics, and scientific literacy \cite{harlen2001assessment}. Tests developed and mandated by the OECD are administered in three-year cycles since its initiation in 2000. The resulting comprehensive dataset offers valuable information for comparing subject performance at continental, regional, national, and sub-national levels over time. The multi-level nature of this dataset makes it suitable for illustrating our proposed visualisation principles for exploratory modelling analysis of BHMs. For the purpose of this analysis, models are fit to a subset of the PISA data using the mathematics average scores for forty(40) European countries from 2003 to 2018. PISA 2000 results in mathematics are not considered, since the first full assessment in mathematics took place in 2003. The 2021 results were delayed to 2022 and were not published until recently, due to the global pandemic.

Our objective is to create visualisations that evaluate model fits, posterior parameter and hyper-parameter distributions and facilitate model comparison. We present five(5) principles for effective visualisation of models in data space and for examining a collection of multiple models simultaneously. We extend the conventional way of displaying model fits on the data points to arrange the individual levels by their respective hierarchical structures, using distinct formatting, layouts, ordering, scaling and colouring. The goal is to enhance the interpretation of model results with respect to information sharing and to explore variations in parameter estimates across different models and hierarchical levels.

The outline of this paper is as follows: Section \ref{background} provides an overview of the background to this study, focusing on the governing essential strategies for visualising statistical models as proposed by \cite{wickham2015visualizing}.
Section \ref{meths} outlines the principles upon which our new visualisation designs are based, using the PISA dataset for illustration, and we discuss insights gained from the PISA analysis, facilitated by the new visualisations. In Section \ref{PISA2022}, we explore the most recent year (2022) of PISA data and compare the observed estimates with model-based predicted estimates. Finally, in Section \ref{dis}, we conclude with an evaluation of the potential benefits and drawbacks of our methods, as well as other related future research.

\section{Background} \label{background}

Visual representations of models offer significant value as complementary tools to numerical summaries. In the Bayesian context, several researchers have investigated the effectiveness of visualisation in developing, computing, validating and evaluating statistical models \cite{gabry2019visualization, gelman2007data, raudenbush2002hierarchical,  correa2023bayesian}.  \cite{wickham2015visualizing} highlighted three important strategies for visualising statistical models: (1) displaying the model in data space,  (2) examining a collection of multiple models, and (3) exploring the process of model fitting. In this work we consider the first two strategies only, which underscore the importance of visualisation in comprehending how a model summarises data, reacts to parameter changes, and fits the data.
 
\begin{figure}[H]
     \centering
     \includegraphics[keepaspectratio]{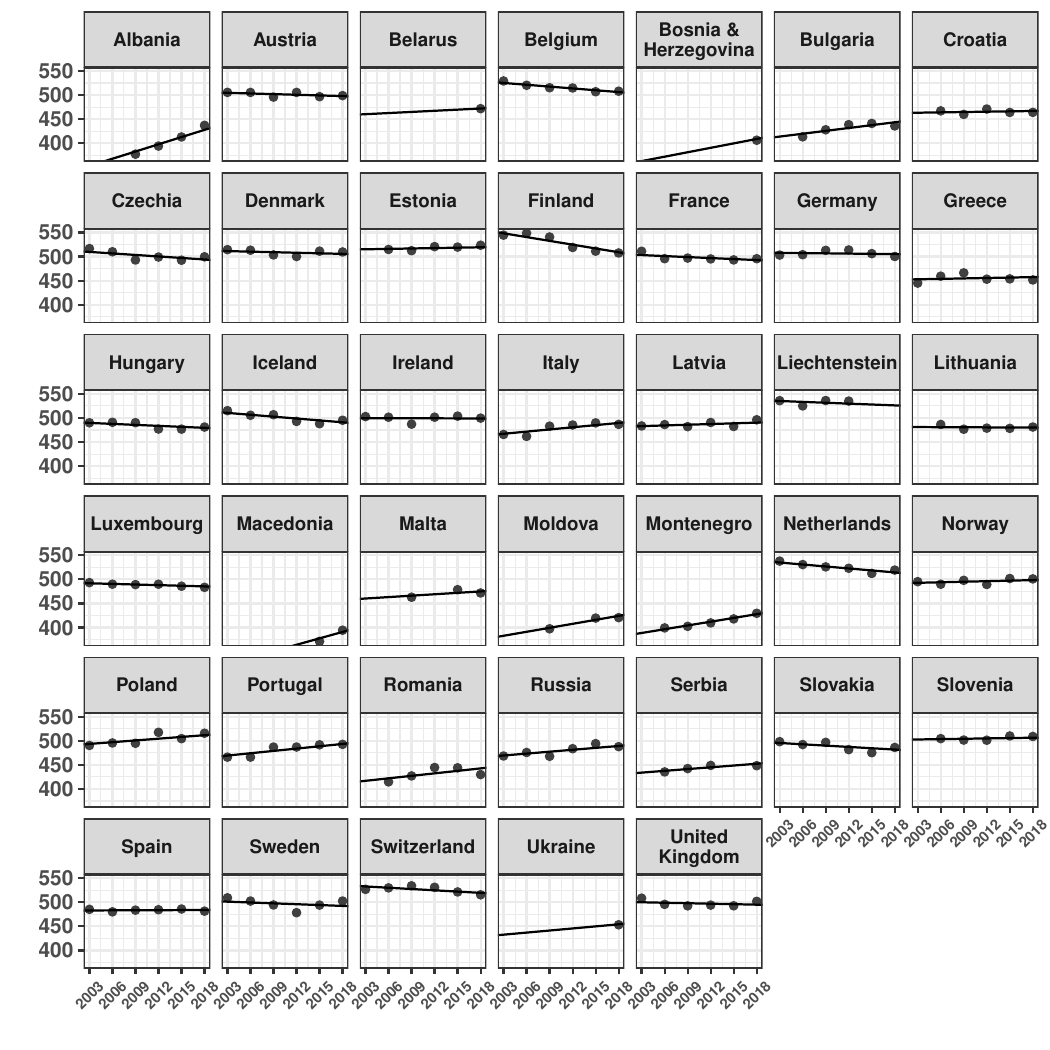}
     \caption{Visualisation of PISA European mathematics scores with model fit, showing overall trends across each country. Belarus, Bosnia \& Herzegovina and Ukraine, each with one data point inherit the common global slope, due to the hierarchical model.}
     \label{Country_specific}
 \end{figure}

There are various conventional approaches for displaying the model in data space. For illustration, Figure \ref{Country_specific} presents a display of estimates from a model fit to mathematics scores from the PISA dataset for 40 European countries using \pkg{brms} \cite{brms}. Here, regression fits are displayed on the observed data, faceted by country, indicating that the linear model broadly captures the patterns in the data and provides information on the direction of trends for each country.

The model equation for the displayed fit is specified in \pkg{brms} as:
\begin{equation}
    \text{math} \sim \; \text{year} + (1 + \text{year} | \text{country}) \tag{country model}
\end{equation}

This model leverages global parameter distributions to enhance parameter estimation at the country level. Therefore, linear fits for countries with one data point (Belarus, Bosnia \& Herzegovina and Ukraine) are obtained from the global parameter estimates, as can be seen in Figure \ref{Country_specific}. With this exception, the display provides no insight into the hierarchical structure of the model.

\begin{figure}[H]
     \includegraphics[keepaspectratio]{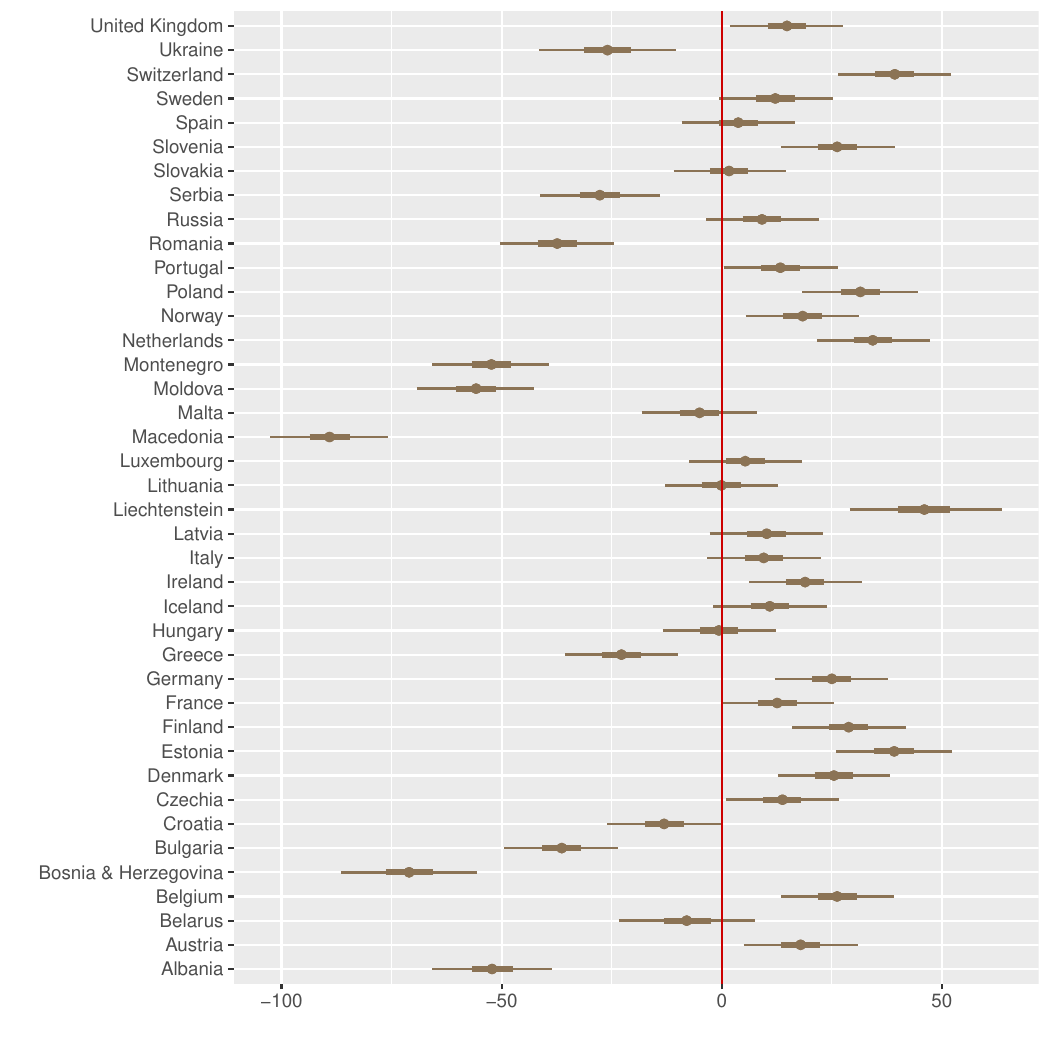}
     \caption{The intercept offset median and 95\% credible intervals for the 40 European countries, showing the deviation from the global estimate. Hungary, Lithuania, Slovakia, and Spain have little deviation from the global estimate, while 2018 PISA mathematics performance for Macedonia and Bosnia \& Herzegovina is comparatively low.}
     \label{offset}
 \end{figure}

In hierarchical modelling, observations are structured into nested levels allowing models to capture heterogeneity and shared characteristics across different levels of the hierarchy. This can result in more stable and robust estimates \cite{gelman2007data}. For example, countries in the PISA data could be grouped by geographical region. Hierarchical modelling also automatically incorporates the shrinkage of parameter estimates (i.e., pulling parameter estimates toward the overall hierarchical distribution) and this shrinkage can reduce the influence of outliers. These characteristics of hierarchical modelling prove particularly valuable in situations where sparse data poses a challenge \cite{rue2007approximate}. Hence, displaying the estimates from these types of models require showing the information sharing between the individual level and the hierarchical structures.

Parameters at a higher level of hierarchical models typically represent global or group-level quantities, and the deviation of the individual level from the global estimates or the group-level (hyper-parameter) estimates are the ``offsets''. Figure \ref{offset} shows parameter offsets of the hierarchical model presented in Figure \ref{Country_specific} illustrating the offsets of each country from the global mean. (Note that our models are parameterised so that the intercept represents mathematics scores in 2018, the last year used in the model fits). This conventional method of displaying offset parameters provides insight into how each individual level moves away from the global estimate but provides no information on the global estimate itself. 

For the second strategy outlined in \cite{wickham2015visualizing}, it is advisable to consider and visually compare various competing models. Consequently, in our context, there is a need for visualisation tools specifically designed to present parameter estimates from multiple hierarchical models. A simple representation in this context might be a display in the style of Figure \ref{offset} but which shows intercept or slope estimates across multiple models side-by-side. However, richer and more informative displays are obtained by incorporating model hierarchies and hyper-parameter estimates.

In this work, we design new visualisations for BHMs that improve on Figures \ref{Country_specific} and  \ref{offset} by incorporating hierarchical information from various levels in the model hierarchy. In the design of these visualisations, there are various principles to consider, for example, those described by \cite{munzner2014visualization} and \cite{wilke2019fundamentals}. Both authors highlight the need for consistency in design choices, such as uniformity of scaling and consistency in the use of colour and formatting. \cite{wilke2019fundamentals} emphasises the use of multiple panels for large and complex datasets, such that the data can be segmented based on the one or more dimensions of the data, visualised separately, then arranged into a grid for an easy interpretation. We have adapted these design principles to explore model estimates in data space and examine a collection of multiple models, highlighting prominent disparities and similarities between and within the hierarchical structures.

\section{Visualisation Principles} \label{meths}

In this section, we present the key principles of our visualisation approach. In Section \ref{M in ds} we apply these principles to the design of a display for BHMs in data space and in Section \ref{M-models}, we further apply these to the design of visualisations for the comparison of multiple BHMs.

To illustrate our visualisation approaches, we employ the previously used subset of the PISA data. We decided to consider country groupings according to geographical regions, as defined by the World Health Organisation \cite{who}, and level of income, using the World Development Indicators, provided by the WDI \cite{wdi} \proglang{R} package. These indicators categorised countries into lower-middle, upper-middle and high-income by their population, environment, economy, global linkages, states and markets. Within the dataset under study, only Ukraine belongs to the lower-middle income category. Consequently, we merged both lower-middle income and upper-middle income classifications into a single ``middle-income'' category, resulting in two distinct income categories: middle income and high income.

We fit various BHMs to the PISA data, using the \pkg{brms} package, which is a front-end for \proglang{Stan} \cite{carpenter2017stan}.

\subsection{Model in data space} \label{M in ds}

Visualising the model in the data space requires the presentation of model estimates alongside observed data. Ideally, this should be done in accordance with the model structure to enhance understanding of the model fit in relation to the model specification. In hierarchical modelling, individual-level observations are nested within higher-level groupings to account for the variability between groups and within groups and to capture the average characteristics or influences of the groups on the individual level. An example for the PISA data is to fit a model using a geographical (country within region) hierarchical structure. The model equation is specified in \pkg{brms} as:
\begin{equation}
      \text{math} \sim \; \text{year} + (1 + \text{year} | \text{region}) + (1 + \text{year} | \text{country})   \tag{region model}
\end{equation}
 
This model uses regional-level distributions to enhance parameter estimation at the country level. The regional parameters are informed by the global mean, while at the country level, parameter estimates are drawn from their respective regional-level distribution.

\begin{figure}[H]
     \centering
     \includegraphics[keepaspectratio]{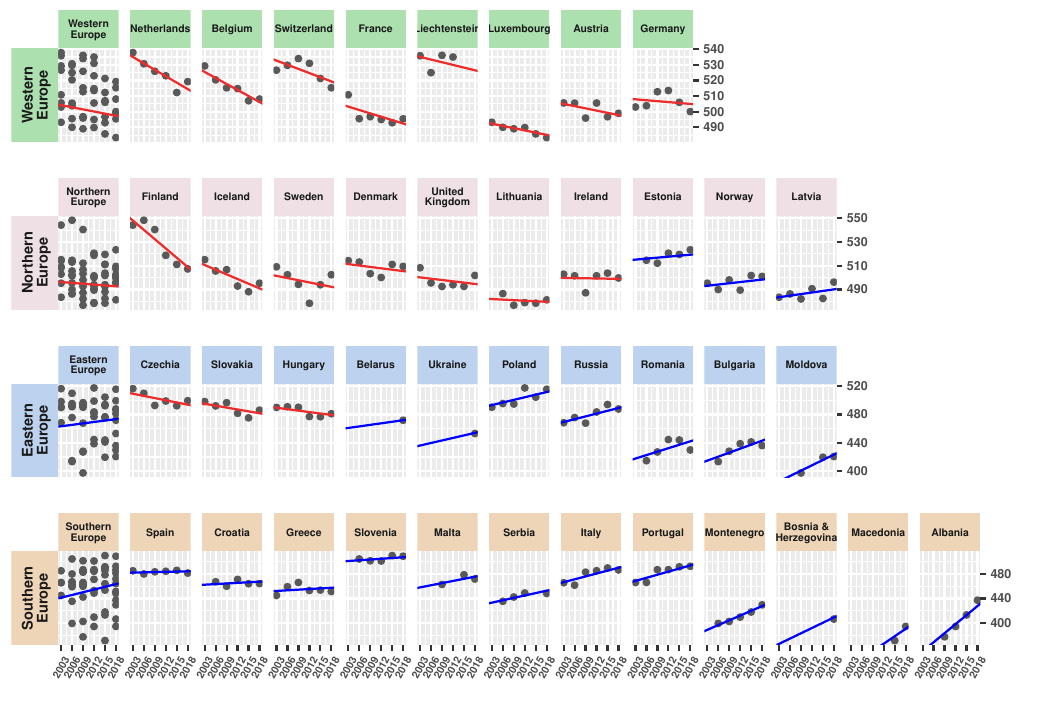}
     \caption{Regional hierarchical model fits to mathematics scores of PISA European data. Western and Southern European countries have a consistent negative and positive trend respectively, but trends are mixed for Northern and Eastern groupings.}
     \label{Region_group}
 \end{figure}

Figure \ref{Region_group} illustrates the regional hierarchical model in data space. This figure improves Figure \ref{Country_specific} (where countries are ordered alphabetically) to order countries according to the regional structures used in the model fitting. Here, panel layouts and colours tailored to the regional structure of the data facilitate comparisons within each region. This approach also displays the regional fit (as determined by the hierarchical model hyper-parameters) superimposed on regional data, in the first column of each row.

We used the following principles in the design of Figure \ref{Region_group}:

\textbf{P1 - One panel per individual-level (unit):} 
In Figure \ref{Region_group}, like Figure \ref{Country_specific}, we use one panel to display the model result in data space per country, as the country level is the lowest level. This clearly reveals the country trends, allowing anomalies to stand out, such as the observation for Ireland in 2009, Liechtenstein in 2006 and Sweden in 2012. In the case of fewer countries or fewer countries per region, visualisation designs that overlay countries could be considered.

\textbf{P2 - Panel grouping to show the model's hierarchical structure:} 
In the realm of hierarchical modelling, our objective is to organise all individual level units according to the hierarchical structure of the model to facilitate comparison between and within groups. In Figure \ref{Region_group}, each region appears in a separate row. Each panel, representing a country, is organised based on its respective geographical region. We see that  mathematics scores in Western Europe are decreasing, while those in Southern Europe are increasing. In addition, this principle facilitates the identification of the mixed trends across countries in Eastern and Northern Europe, despite overall positive and negative trends in the Eastern and Northern European regions respectively.

\textbf{P3 - Colour to differentiate groups and to enable comparison:} 
In Figure \ref{Region_group}, colours are used in two ways. Within each panel, to distinctly differentiate the direction of the model fit on observed data, negative trends are shown in {\color{firebrick2}red} and positive trends in {\color{blue}{blue}}. Colour is also used alongside position to group countries, with labels for Western Europe in the {\color{lightgreen}light-green} colour, Northern Europe with {\color{lavenderblush3}lavender}, Eastern Europe in {\color{steelblue}steel-blue}, and Southern Europe in the {\color{bisque2} bisque} colour. The same region colours will be used consistently in other displays.

\textbf{P4 - Panel ordering to enable model comparisons:}
In Figure \ref{Region_group}, regions (rows) are ordered by increasing regional slope parameter estimates. Furthermore, within each region, countries are arranged by increasing slope. This arrangement places Finland first in the Northern Europe group because, while it has high mathematics scores, these scores dropped rapidly over time. Belarus and Ukraine are adjacent because, with one observation apiece, the slope displayed in their panels is the overall Eastern Europe slope (slope hyper-parameter estimate for the region).

\textbf{P5 - Panel scaling for effective comparisons:} 
Ideally, in displays such as Figure \ref{Region_group}, panels would use the same scaling to facilitate comparison across countries. However, as the range of differences in Western and Northern Europe is small, this choice obscures trends within each country as most slopes appear flat in these groups. Rather than free scaling across all panels, in this case we have opted to use the same scaling for the panels in each row and different scaling across rows. This choice enables easy comparison of countries within each region. However, comparison of mathematics scores across countries in different regions requires careful inspection of the row axes.

\subsection{Examining a collection of multiple models} \label{M-models}

Displaying the model fit by region, as seen in Figure \ref{Region_group}, noticeable patterns (positive and negative trends) emerge in Eastern and Northern Europe. To identify additional sources of variation in the dataset, we further classified countries based on their respective geographical region and income level and introduced another grouping structure termed ``income-region'', which combines region and income as a new variable, leading to three identified sources of heterogeneity and shared characteristics (region, income and income-region) to construct hierarchical models.

For comparison purposes, we consider five(5) linear models fit to the PISA data:
The first three models are a non-pooled model, a country-specific hierarchical model where country grouping provides an additional source of variation, 
and a two-level hierarchical model where countries are nested in region and both region and country are sources of variation.
The non-pooled model assumes all observations are independent, while the hierarchical models leverage group-level distributions to enhance parameter estimation at the country level. These three models are specified in \pkg{brms} as:
\begin{align} 
    &\text{math} \sim \; \text{year} * \text{country} \tag{1: non-pooled} \\
    &\text{math} \sim \; \text{year} + (1 + \text{year} | \text{country}) \tag{2: country model}\\
    &\text{math} \sim \; \text{year} + (1 + \text{year} | \text{region}) + (1 + \text{year} | \text{country}) \tag{3: region model}
\end{align}
The equations for two further models (4) and (5) are the same as (3), except with the grouping variable region replaced by income and income-region respectively. Note that model 2 appeared previously in Figures \ref{Country_specific} and \ref{offset}, while model 3 was displayed in Figure \ref{Region_group}.

\begin{figure}[H]
     \centering
     \includegraphics[scale = 0.45]{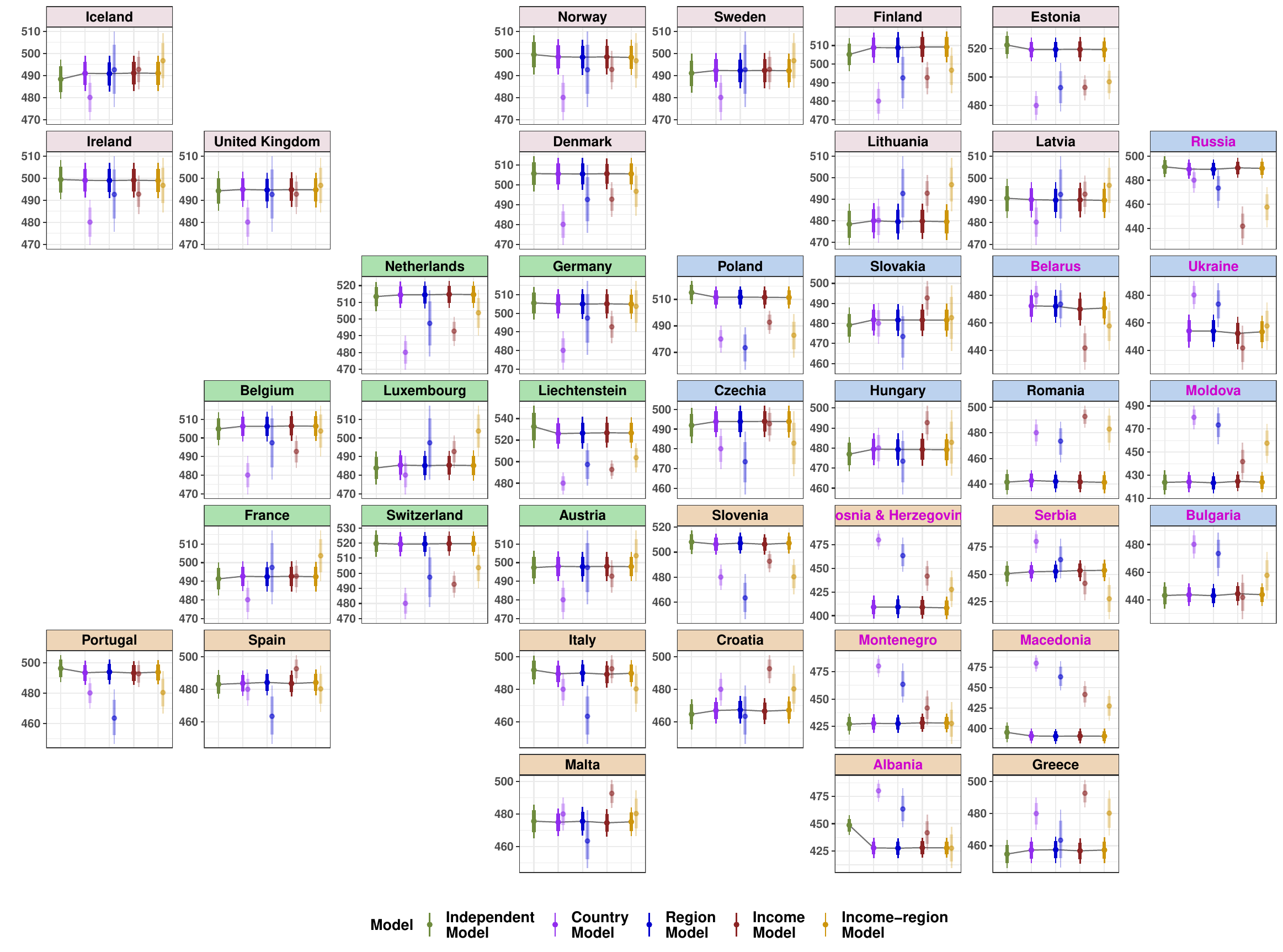}
     \caption{Intercept distributions (2018 median estimates with 80\% and 95\% credible intervals) from five model fits to mathematics scores from European PISA data. Hyper-parameter estimates are presented in lighter shades beside the parameter estimates. Countries are arranged according to the map layouts and labels are coloured to differentiate the income categories. Panels are individually scaled, because of the variation in estimates across countries.}
     \label{geo.intercept}
 \end{figure}

\begin{figure}[H]
     \centering
     \includegraphics[scale= .45, keepaspectratio]{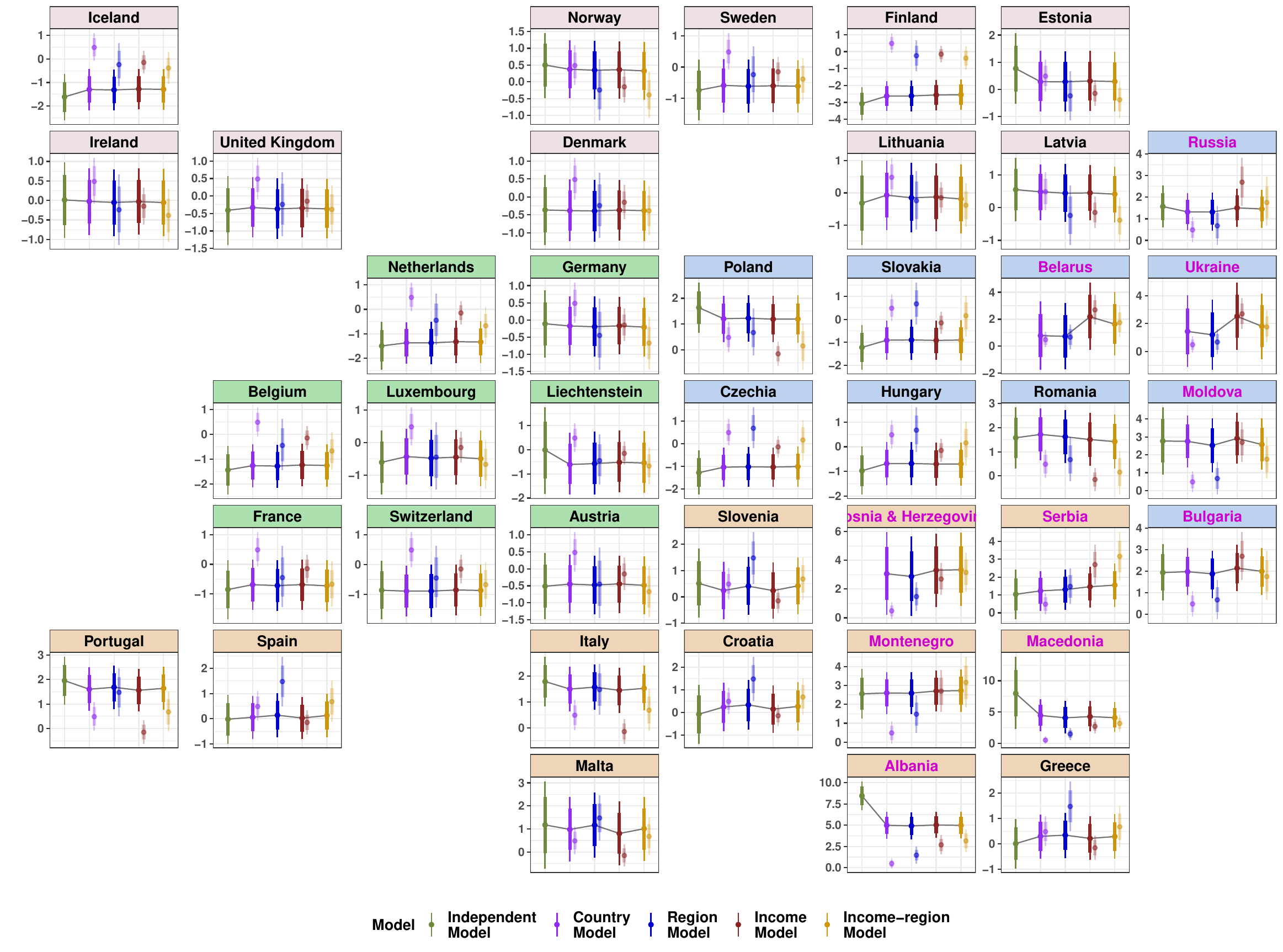}
     \caption{Slope distributions (median estimates with 80\% and 95\% credible intervals) from five model fits to mathematics scores from European PISA data. Hyper-parameter estimates are presented in lighter shades beside the parameter estimates. Countries are arranged according to the map layouts and labels are coloured to differentiate the income categories. Panels are individually scaled, because of the variation in estimates across countries.}
     \label{geo.slope}
 \end{figure}

We chose the intercept and slope parameters to summarise and interpret the model results while examining a collection of multiple models simultaneously. Recall, our models are parameterised so that the intercept represents mathematics scores in 2018. In Figure \ref{geo.intercept}, we display the median and 80\%, 95\% credible intervals of the intercept for each country alongside their respective hyper-parameter distributions from models (1) to (5). Similarly, Figure \ref{geo.slope} presents the slope estimates. Comparing the estimates from various models highlights the shrinkage effects.

In Figure \ref{geo.intercept} and Figure \ref{geo.slope}, each model is represented by a vertical axis, with country-level estimates shown in darker shades and hierarchical (hyper-parameter) distributions in lighter shades alongside. Countries are represented by separate panels, carefully arranged by the map layout following the regional structure, with labels coloured to differentiate income categories. The panels are individually scaled due to the variation in estimates across countries, facilitating comparison of models within each country.

The principles P1--P5 were also used in the design of our Figures \ref{geo.intercept} and \ref{geo.slope}:\\
\textbf{P1 - One panel per individual-level (unit):} 
As in Figure \ref{Region_group}, one panel per country is used in Figures \ref{geo.intercept} and \ref{geo.slope}. Hence, each panel represents the country intercept/slope parameter distribution respectively using median, 80\% and 95\% credible intervals as distribution summaries. Models are arranged along the horizontal axis, in order of model (1) to (5). The country estimates are connected using a horizontal dashed line to differentiate them from the estimates from the higher-level groups.

\textbf{P2 - Panel layout to show the model's hierarchical structure:} 
Rather than arranging the countries by region as in Figure \ref{Region_group}, we use a rough geographical map layout to arrange country panels, so that countries in the same region group are positioned nearby. 

\textbf{P3 - Colour to differentiate groups and to enable comparison:} 
Figures \ref{geo.intercept} and \ref{geo.slope} use colours in a number of ways: (1) to differentiate region groups and income groups, (2) to distinguish each model fit. The same region colour scheme used in Figure \ref{Region_group} was employed and colour is further used for the country label to indicate the income level, high income in black and middle income in {\color{magenta}magenta}. Note that the middle income countries are mostly grouped together on the right hand side. In addition, we use one colour per model fit. Within each panel of Figures \ref{geo.intercept} and \ref{geo.slope}, the hierarchical structure parameter estimates (hyper-parameter) for the hierarchical models (2) to (5) are shown in a lighter shade positioned beside each model's parameter estimates. For model (2), the global intercept and slope distributions are in {\color{lightpurple} lightpurple}, while {\color{lightblue}lightblue}, {\color{brown4}brown} and  {\color{darkgoldenrod3}darkgoldenrod} are similarly used for models 3, 4 and 5.

\textbf{P4 - Panel ordering to enable model comparisons:}
Within the panels of Figures \ref{geo.intercept} and \ref{geo.slope}, models are ordered from (1) to (5), going from the simplest non-pooled model (1), followed by the country-specific one-level hierarchical model (2), then the region hierarchical model (3) with two levels (country in region), and models (4) and (5) also with two levels (country in income and country in income-region) respectively.

\textbf{P5 - Panel scaling for effective comparisons:} 
In Figure \ref{geo.intercept}, our main goal is to compare the intercept distributions from the five models within each country. Similarly, Figure \ref{geo.slope} for the slopes. Therefore, we use free scaling across all panels, so that differences across the distributions for the models are visible in each of the countries. 

The principles 1--5 underlying our visualisations are straightforward and aim to visualise model estimates that align closely with the model specification. Ideally, this means being able to compare estimates both within and across various hierarchical structures and to interpret these estimates while considering the effect of information sharing.

\subsection{Insights from PISA model visualisations} \label{insights}

Here, we will highlight key insights from the visualisations of PISA models in Figures \ref{Region_group}, \ref{geo.intercept} and \ref{geo.slope}.

The structured arrangement of Figure \ref{Region_group} (showing model (3)), and the use of colours to indicate the direction of trends, promotes pattern recognition across countries, such that similar trends and disparities can be identified within the regional groups that govern the hierarchical structure of the model. For instance, the Eastern European countries Czechia, Slovakia and Hungary exhibit negative trends despite the overall positive trend in the region. Likewise in Northern Europe, Estonia, Norway and Latvia demonstrate positive trends over time notwithstanding the negative regional trend. 

Within each row of Figure \ref{Region_group}, countries are ordered by increasing slope, so we can easily identify countries with the most and the least decline in the region. In Western Europe, Netherlands and Belgium have a high rate of decline. Finland had the highest PISA score in 2003, but has the steepest decline across Northern European countries and indeed overall, though this last assessment is made more difficult by the different vertical axis scaling across the rows. Similarly, Albania is notable for its low initial PISA scores and by far the highest rate of improvement across the years.

Next we move to Figures \ref{geo.intercept} and \ref{geo.slope}, which compare five different PISA models.
For all countries, there is little variation in the country intercept estimates (corresponding to PISA performance in 2018; Figure \ref{geo.intercept}) across the four hierarchical models (2--5), even though the estimates for the groups depend on the choice of hierarchical model (see for example the group estimates for the Southern Europe middle income countries, whose labels have {\color{magenta}magenta} text and {\color{bisque2} bisque} background). However, comparing  the non-pooled model (1), we see that for Albania pooling produces a notable reduction in the intercept estimate, while for Finland and Iceland, pooling produces a small increase in the intercept estimate.

Looking at the slopes in Figure \ref{geo.slope}, pooling produces a notable reduction in the slope estimate for Albania and Macedonia, and small increases for Finland and Iceland. The non-pooled fit produces a practically zero slope estimate for Liechtenstein, while the hierarchical models estimate a reduction of about 0.5 per year. For most countries, slope estimates and credible intervals are not affected much by the choice of hierarchical model. There are some small differences in slope estimates evident for the middle income Eastern European countries Belarus and Ukraine, with the income model in {{\color{brown4}brown}} giving higher slope estimates. This discrepancy is not surprising as these countries have just one observation apiece. For the other country with just one observation Bosnia \& Herzegovina, the slope estimates are remarkably stable across models, though with a high-level of uncertainty across all models.

\section{Investigating PISA 2022 Predictions} \label{PISA2022}

Recall that all model results displayed previously were based on PISA results up to 2018.
It will be interesting to compare model-based predictions for 2022 with the the observed data available from the most recent PISA survey in 2022 \footnote{The PISA test was originally due to be conducted in the year 2021 but was delayed by one year because of the COVID-19 pandemic. The exceptional circumstances throughout this period, including lock-downs and school closures in many countries, which led to the difficulties in collecting some data \cite{organisation2023pisa}.}. A comparison of models 2--5 via leave-one-out cross-validation \cite[LOO-CV][]{LOO} shows little difference between models, agreeing with our findings from the visualisations of Figures \ref{geo.intercept} and \ref{geo.slope}. Therefore, we chose the richest hierarchical model for this investigation, which is the income-region model (5). In ideal circumstances this comparison could be used to assess the model’s predictive accuracy. However, as we anticipate PISA results have been negatively impacted by COVID-19, this comparison will instead allow us to compare the level of impact across countries. 

 \begin{figure}[H]
     \includegraphics[keepaspectratio]{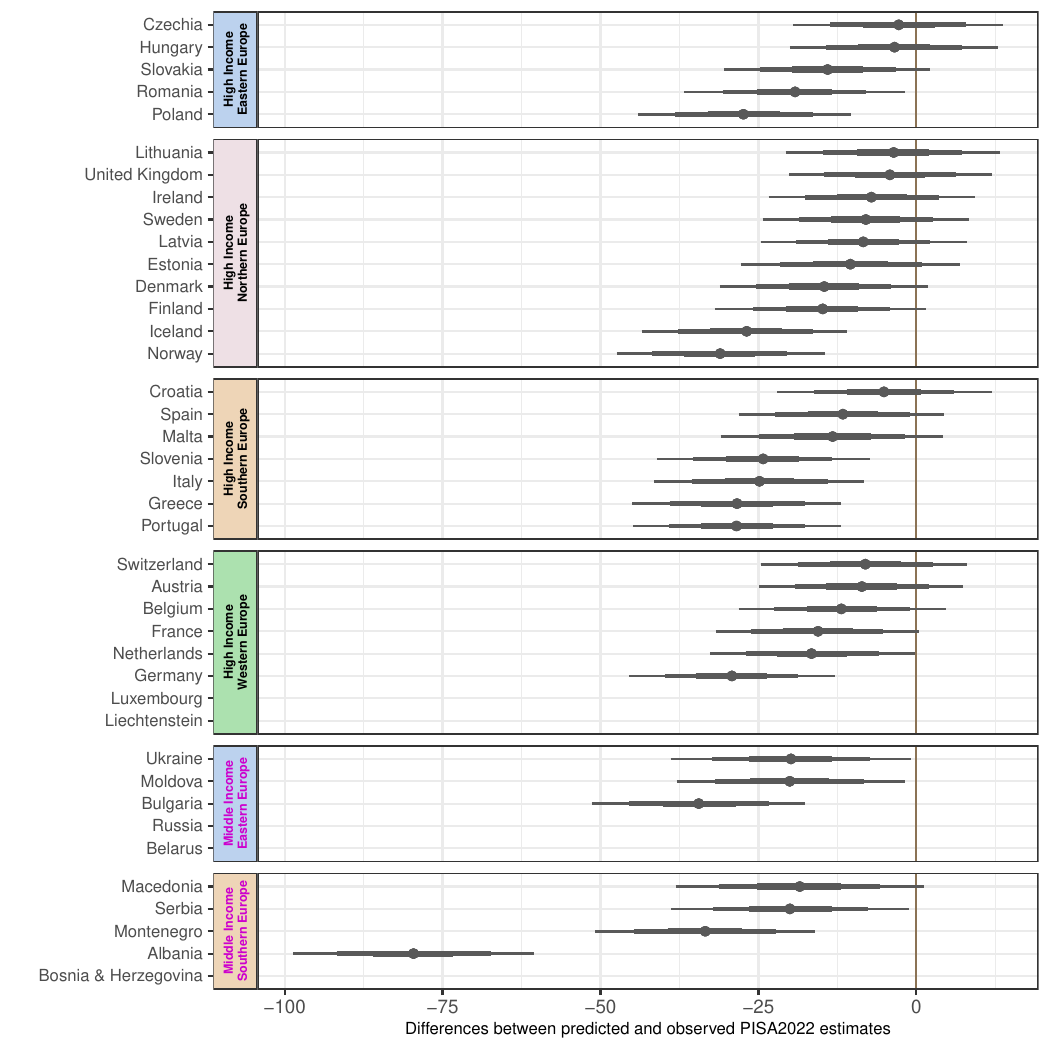}
     \caption{Differences between observed PISA maths data for 2022 and  predictions from the income-region model (5), alongside 80\%, and 95\% credible intervals. Note that Luxembourg, Liechtenstein, Belarus, Russia, and Bosnia \& Herzegovina did not participate in the PISA 2022 survey.}
     \label{Diff_IRM}
 \end{figure}
 
In Figure \ref{Diff_IRM}, we present the prediction error (observed - predicted) with the 50\%, 80\%, and 95\% prediction intervals for the 35 out of 40 European countries that participated in the PISA 2022 survey. We arrange countries by their respective income-region group, order countries within each group according to their increasing prediction error and utilise the same region and income colour schemes as proposed in Section \ref{meths} to distinctly label the income-region groups.

In all countries the PISA 2022 values are lower than predicted. The country with by far the biggest drop is Albania, and in fact this 2022 score has wiped out all of the gains achieved since 2009. A possible reason for this decline, in addition to COVID-19, is that Albania experienced a strong earthquake in 2019 which caused damage to infrastructure, including schools and houses. It is also evident that PISA mathematics scores for middle income countries showed higher decline than those in high-income countries, suggesting that high-income countries were in a better position to mitigate the effect of COVID-19 on the school system.

\section{Discussion} \label{dis}

We have presented new visualisations and proposed informative principles for displaying hierarchical model estimates according to the strategies outlined by \cite{wickham2015visualizing}. The principles in Section \ref{M in ds} focused on the display of model results in the data space to provide an in-depth examination of the model fit in relation to the hierarchical structure of the model. We utilised layout, scales, colours, point sizes, and line types, to differentiate and emphasise patterns and to facilitate comparisons. The same principles (Section \ref{M-models}) offer a systematic approach for examining a collection of multiple models, facilitating comparisons of parameter and hyper-parameter distributions across models with different hierarchical structures. 

In Section \ref{insights}, we demonstrated that the new visualisations led us to identify countries with similar performance and those with unusual performance in PISA mathematics results. In this case study, the choice of hierarchical model is not crucial. For most countries, there is little to choose between models 2--5,  except for a few countries with very little data.
Applying the same visualisation principles in Section \ref{PISA2022} to compare PISA 2022 mathematics results with model predictions established that across all countries, 2022 results were lower than expected (likely due to COVID-19), drastically so in the case of Albania, though overall, high income countries were less impacted.

The model fits and summaries presented in this paper utilised the following \proglang{R} packages, \pkg{brms} \cite{brms}, \pkg{dplyr} \cite{dplyr}, and \pkg{tidybayes} \cite{tidybayes}.
The visualisations were obtained using the following \proglang{R} packages, \pkg{ggplot2} \cite{ggplot2}, \pkg{ggragged} \cite{ggragged}, \pkg{ggdist} \cite{ggdist}, and \pkg{geofacet} \cite{geofacet}, all available from the Comprehensive R Archive Network (CRAN) at \url{https://cran.r-project.org/}. 
The \pkg{ggragged} \proglang{R} package is designed for creating arrays where rows can have different lengths and can be manipulated to suit desired arrangements, as in Figure \ref{Region_group}. The \pkg{ggdist} \proglang{R} package is designed to present distributions and uncertainty, as seen in Figures \ref{geo.intercept} and \ref{geo.slope}. The package displays the median intercept and slope respectively along with their credible intervals using point intervals. 
The \pkg{geofacet} \proglang{R} package is designed for creating spatial visuals in a clear and comprehensible manner by arranging panels to mimic the geographical map layout, as in Figures \ref{geo.intercept} and \ref{geo.slope}. 

The new visualisation designs are simple, flexible, adaptable and can be applied to non-Bayesian models, to models that fit relationships beyond linear regression, and to models with a richer hierarchical structure. In our illustrative example there are 40 countries under study, having a larger number of individual levels might result in clutter, which could be mitigated by selecting a smaller number of individual units for detailed examination. The visualisations presented used layouts based on spatial positioning of countries in Figures \ref{geo.intercept} and \ref{geo.slope}. However in other applications, these model comparison visualisations could use layouts based on groups such as those in Figure \ref{Region_group}. We acknowledge that displaying hierarchical structures richer and more complex than those illustrated here could be challenging, but the principles described in \ref{M-models} should provide helpful guidance.

For future work, designing a visualisation for the joint distribution of the intercept and slope estimates in the context of hierarchical models would be a useful next step. Additionally, providing an \proglang{R} package that allows users to select relevant parameters to be visualised from the model will also be a great addition.  

In conclusion, our principles for visualising hierarchical model estimates provide a comprehensive and insightful approach to understanding complex model structures. By following our step-by-step principles, researchers can achieve a clearer understanding of hierarchical structures present in the data, and communicate results more effectively. Despite potential challenges, such as managing large numbers of individual levels and hierarchical structures, our approach remains highly valuable for examining and comparing hierarchical models. Through these visualisation principles, we contribute to advancing the field of hierarchical modelling, offering tools and insights that can be adapted to a wide range of research contexts.

\section*{Data and Code Availability}

The source code and data for reproducing the results in this paper is available at \url{https://github.com/Oluwayomi-Project/Visualisation-for-exploratory-analysis-of-BHM}.

\newpage
\section*{Acknowledgments}

``This publication has emanated from research conducted with the financial support of Science Foundation Ireland under Grant number 18/CRT/6049''. For the purpose of Open Access, the author has applied a CC BY public copyright licence to any Author Accepted Manuscript version arising from this submission.
\newpage
\bibliographystyle{unsrt}  
\bibliography{references}

\end{document}